\documentclass[showpacs,showpacs,superscriptaddress]{revtex4}
\usepackage{graphicx}
\usepackage{dcolumn}
\usepackage{amsmath}

\makeatletter
\def\btt#1{\texttt{\@backslashchar#1}}%
\DeclareRobustCommand\bblash{\btt{\@backslashchar}}
\makeatother

\begin{document}


\title{Aging in humid granular media}
\author{Fr\'ed\'eric Restagno}
\thanks{To whom correspondence should be addressed.
Now at the Laboratoire des fluides organis\'es, Coll\`ege de
France, 11 place Marcelin Berthelot, 75005 Paris, France. E-mail:
frederic.restagno@college-de-france.fr} \affiliation{Laboratoire
de Physique - \'Ecole Normale Sup\'erieure de Lyon - 46 all\'ee
d'Italie, 69364 Lyon cedex 07, France} \affiliation{D\'epartement
de Physique des Mat\'eriaux - Universit\'e Claude Bernard - Bat.
L\'eon Brillouin, 43 bd du 11 Novembre 1918, 69622 Villeurbanne
cedex, France}
\author{C\'ecile Ursini}
\affiliation{Laboratoire de Physique - \'Ecole Normale
Sup\'erieure de Lyon - 46 all\'ee d'Italie, 69364 Lyon cedex 07,
France}
\author{Herv\'e Gayvallet}
\affiliation{Laboratoire de Physique - \'Ecole Normale
Sup\'erieure de Lyon - 46 all\'ee d'Italie, 69364 Lyon cedex 07,
France}
\author{\'Elisabeth Charlaix}
\affiliation{D\'epartement de Physique des Mat\'eriaux -
Universit\'e Claude Bernard - Bat. L\'eon Brillouin, 43 bd du 11
Novembre 1918, 69622 Villeurbanne cedex, France}

\date{\today}

\begin{abstract}
Aging behaviour is an important effect in the friction properties
of solid surfaces. In this paper we investigate the temporal
evolution of the static properties of a granular medium by
studying the aging over time of the maximum stability angle of
submillimetric glass beads. We report the effect of several
parameters on these aging properties, such as the wear on the
beads, the stress during the resting period, and the humidity
content of the atmosphere. Aging effects in an ethanol atmosphere
are also studied. These experimental results are discussed at the
end of the paper.

\end{abstract}


\maketitle

\section{Introduction}
Granular media have many interesting and unusual properties
\cite{Jaeger96}.  The intrinsic dissipative nature of the
interactions between the constituent macroscopic particles sets
granular matter apart from conventional gases, liquids or solids.
One of the most interesting phenomena in granular systems is the
transition from a static equilibrium to a granular flow. Contrary
to ordinary fluids, they can remain static even with an inclined
free surface. But when the angle of the surface with the
horizontal exceeds some threshold value $\theta_m$, the pile
cannot sustain the steep surface and starts to flow until its
angle relaxes under a given angle of repose $\theta_r$.

The friction properties of grains play an important role in this
transition from static equilibrium to flow \cite{Degennes99}.
Indeed, the phenomenological laws of static friction formulated by
Coulomb in 1773 identified
the existence of a definite avalanche angle in a 
granular media to a friction coefficient: $\tan \theta_m = \mu$.
Subsequent works have emphasised the similarities between
solid-solid friction and friction in granular media. On the one
hand, detailed studies in solid-solid friction have brought into
evidence temporal evolution in friction phenomena
 between solids.  In a number of materials the static friction coefficient
increases logarithmically with the contact time, whereas the
dynamic friction coefficient decreases as the logarithm of the
slipping velocity\cite{Rabinowicz}. These temporal evolutions have
been shown to be an effect of the evolution over time of contacts
between surfaces asperities - the aging of contacts - , and result
in stick-slip behaviour at low velocity motion \cite{Heslot94}.
Recently, Nasuno et al. \cite{Nasuno98,Geminard99} have reported
measurements of stick-slip transition on granular layers of
spherical glass beads. A similar study has been made by Lubert et
al. \cite{Lubert2001} between silicagel grains in an annular
geometry. On the other hand, the temporal evolution of friction
properties in a granular material has been studied by Bocquet et
al. \cite{Bocquet98a} , who report a slow increase of the maximum
stability angle of submillimetric glass beads with the time of
rest of the granular heap. Nevertheless, this temporal evolution
is observed only in an humid atmosphere. This important effect of
humidity in the aging properties of friction has also been
reported by  Dieterich in rock-to-rock friction
\cite{Dieterich84}, and Crassous et al. on paper-to-paper and
glass-to-glass friction \cite{Crassous99}. This shows that a small
amount of liquid in the atmosphere can drastically modify the
friction properties of solids.

A number of recent experimental studies have studied the effect of
a small amount of liquid on the maximum stability angle in a
granular material.Barab\'asi et al.
\cite{Hornbaker97,Albert97,Tegzes99,Barabasi99} as well as Mason
et al. \cite{Mason99,Halsey98} have studied the influence of a
small quantity of non-volatile liquid added to a granular medium,
and shown an important increase of its maximum stability angle.
Using water vapor and alcane vapor, Fraysse et al.
\cite{Fraysse99} have reported a significant increase in  the
maximum stability angle with the vapor content of the surrounding
atmosphere.

These experiments do not explicitly incorporate time as a
parameter for cohesion  effects in granular media as studied by
Bocquet et al. \cite{Bocquet98a}. More recently, Ovarlez et al.
\cite{Ovarlez2001} have shown an humidity-induced aging effect on
the dynamical behaviour of a granular column pushed vertically. A
similar effect has been reported by D'Anna et al.
\cite{Danna2000b} in frictional properties of granular in a
Couette geometry. In this paper, we investigate more precisely the
temporal effects on the avalanche angle of submillimetric glass
beads. For this purpose, we use a rotating drum and perform
experiments at a controlled temperature and humidity. We have also
performed experiments in ethanol vapor. We analyse in detail the
effect of parameters such as the tangential stress during the
resting period and the wear of the beads on the temporal
dependence of the avalanche angle. We show that this temporal
evolution results from the increase of the cohesion force between
grains when the beads are in a vapor atmosphere.

\section{Experimental system}

\subsection{The setup}

\begin{figure}[htbp]
\includegraphics[width=12cm]{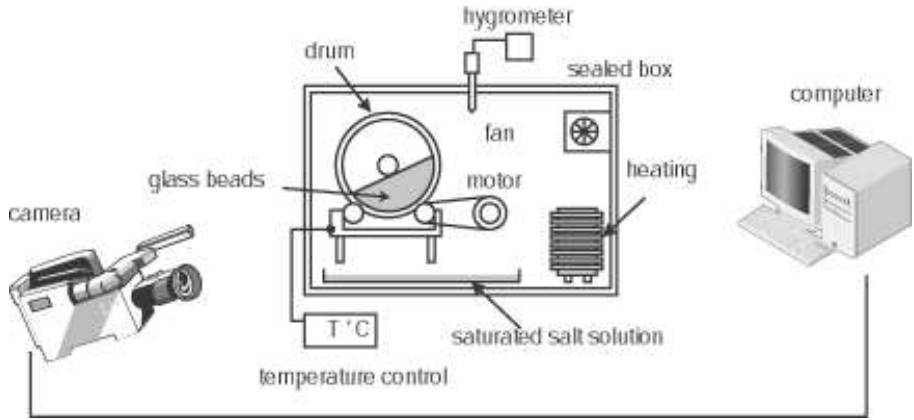}
\caption{Experimental setup.} \label{fig_montage}
\end{figure}

The experimental setup is described in figure \ref{fig_montage}.
The granular material is inserted in a cylindrical drum of
stainless steel with an inner diameter and a length of $10$~cm.
This geometry reduces the influence of the wall on the stability
angle (the typical length scale on which the wall affects the
stability angle is on the order of a few tenths of bead diameters
\cite{Grasselli97}). The lateral faces of the cylinder are made of
glass with an opening at the center sufficiently large to allow
vapor exchange with the outer atmosphere. The drum is partially
filled with glass beads. In all the experiments we have used the
same height of glass beads (4 cm). In order to keep the relative
humidity of the atmosphere constant during the experiments, the
setup is placed in a sealed lucite box. The temperature in the box
is controlled by a heating resistor, a fan enclosed in the box and
a temperature controller. The heating resistor is a THERMOCOAX
heating element connected to a digital temperature controller (REX
F400, TCSA Dardilly, France). The temperature measurement is done
with a Pt100 inserted in the box. The temperature is fixed at $29
\pm 0.2~^{\textrm o}$C.

The cylinder can rotate around its horizontal axis at a constant
speed $\Omega$ controlled with a DC-motor. The rotation speed can
be varied from about 0.01 to 60 rpm. One of the lateral faces of
the cylinder is monitored by a video camera connected to a
computer. The maximum slope of the surface is then directly
measured with SCION Imaging Software. The typical error on the
measurements is typically $\pm 1^{\textrm o}$. The main  error is
due to the lack of flatness of the heap surface. This error is
smaller than the dispersion of the measurements, which are due to
small differences in the preparation of the samples, small drifts,
and unwanted vibrations\dots We report the avalanche angle
measured on the surface of the glass windows on the lateral faces
of the cylinder. This angle is always greater than the avalanche
angle in the middle of the drum. The difference between this two
angles (typically 2°) does not change with the waiting time.

\begin{figure}[htbp]
\includegraphics[width=6cm]{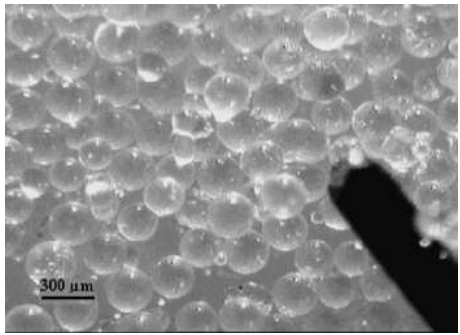}
\caption{Picture of a sample of glass beads with an optical
microscope.} \label{fig_bille}
\end{figure}
The glass beads used in this study are industrially used for
sandblasting (see figure \ref{fig_bille}) and are sold by Matrasur
(Marcoussis, France). These glass beads are smooth and have  a
roughly spherical geometry.

In all the forthcoming experiments, the diameter of the beads is
smaller than 0.3~mm. We have found that temporal effects become
small for glass beads larger than 0.3~mm. We used three samples of
glass beads (hereafter $A, B, C$), with sizes calibrated by
sifting (see table \ref{sample_tab}).

\begin{table}[htbp]
  \centering
\begin{tabular}{|c|c|c|}
\hline
  Sample & Min size ($\mu$m) & Max size ($\mu$m) \\ \hline \hline
  A & 200 & 250 \\
  B & 125 & 160 \\
  C & 0 & 50 \\ \hline
\end{tabular}
\caption{Size of the beads in the different
samples.}\label{sample_tab}
\end{table}

\subsection{Humidity control}

\begin{table}[htbp]
\centering
\begin{tabular}[t]{|l|c|c|c|c|c|c|c|c|c|c|}
\hline
Saturated salt solution & \multicolumn{10}{c|}{$RH\%$}\\
\cline{2-11}
Temperature & 5 & 10 & 15 & 20 & 25 & 30 & 35 & 40 & 50 & 60 \\
\hline \hline Lithium Chloride (LiCl) & 14 & 14 & 13 & 12 & 12 &
12 & 12 & 11 & 11
& 10\\
Potassium Acetate (KCH$_3$CO$_2$) & - & 21 & 21 & 22 & 22 & 22 &
21
& 20 & - & -\\
Magnesium Chloride (MgCl$_2$)& 35 & 34 & 34 & 33 & 33 & 33 & 33 &
32
& 31 & 30\\
Potassium Carbonate (K$_2$CO$_3$)& - & 47 & 44 & 44 & 43 & 43 & 43
&
42 & - & 36 \\
Sodium Chloride (NaCl) & 76 & 76 & 76 & 76 & 75 & 75 & 75 & 75 &
75 & 76 \\ \hline
  \end{tabular}
  \caption{Relative humidity of the air surrounding a saturated salt
solution for different salts at different temperatures.
   (after the french norm NF X 15.014
\cite{Cretinon92}).}\label{salt_tab}
\end{table}
Two parameters are used to quantify the amount of vapor water in
the atmosphere. These are the partial pressure of water $p_{vap}$,
or the relative humidity $RH=p_{vap}/p_{sat}$ , defined as the
partial pressure of water divided by  the saturating  vapor
pressure $p_{sat}$. The relative humidity $RH$ is often expressed
as a percentage: $RH\%=RH\times 100$. The relative humidity is a
more pertinent parameter since it is related to the
thermodynamical properties of the liquid-vapor equilibrium.
Indeed, the undersaturation $\Delta\mu=\mu_{vap}-\mu_{sat}$,
defined as the difference between the chemical potential of the
vapor $\mu_{vap}$ and the chemical potential at the liquid-vapor
equilibrium $\mu_{sat}$, can be expressed as \cite{Defay51}:

\begin{equation}
\Delta\mu=\mu_{vap}-\mu_{sat}=kT\ln\left(\frac{p_{vap}}{p_{sat}}\right
)=kT\ln\left(RH\right)
\end{equation}
where $k$ is the Boltzmann's constant and $T$ the absolute
temperature.

In the literature, different methods for the control of the
relative humidity  are described. Bocquet et al. \cite{Bocquet98a}
used the method of the controlled leak between a water container
and the chamber containing the experiment. This method requires an
adjustable control of the flow of vapor between the two containers
and an accurate hygrometer. Fraysse et al. \cite{Fraysse99} used a
more complicated method consisting in: i) putting the beads under
vacuum, and ii)  injecting a controlled amount of water
(undersaturated vapor). This method is a versatile method that can
be implemented with any volatile liquid, resulting in a shorter
equilibrium time. In our experiments, we used another method to
control the atmosphere. The relative humidity is kept constant by
using an aqueous solution saturated with inorganic salts. A large
beaker of saturated salt solution is put in the box. Each salt
fixes the relative humidity of the surrounding atmosphere at a
given temperature. The purity of the used salt exceeds $99\%$ (in
fact, a small amount of impurity does not induce a significant
drift of the humidity). Table \ref{salt_tab} shows that
temperature has little influence on the relative humidity. The
relative humidity $RH$ is measured during the experiments by a
capacitive hygrometer with a precision of $\pm 2 \%$.
\subsection{Procedure}
Our purpose is to study the aging effects on the avalanche angle
of the granular medium. Therefore, we measure the avalanche angle
of the granular heap after a waiting time $t_w$, the waiting time
being defined as the time elapsed between the onset of an
avalanche and the last overall motion of these grains before the
avalanche. The procedure for measuring the avalanche angle  after
a resting period of length $t_w$ is the following:
\begin{enumerate}
\item First of all, the beads are introduced in the drum and
rotated at a low speed (0.01~rpm) for 48 hours in the controlled
environment. After this preparation, we estimate that the humidity
and temperature are homogeneous throughout the bulk of the
granular medium. \item The drum is rotated for about one minute at
an angular velocity of about 20 rpm, in order to create an overall
motion in the granular medium. \item At time $t=0$ this quick
rotation is stopped. We check that the free surface of the heap
obtained is flat enough so that the value of the maximum slope can
be measured within $2^{\textrm o}$. The position of the drum is
adjusted so that the angle of the free surface of the heap with a
horizontal plane has a well-defined value. We call this angle the
reference angle $\theta_{ref}$. The reference angle is always
smaller than the maximum stability angle of the heap. \item The
granular heap stays at rest until $t\preceq t_w$. The drum is then
set in rotation at very low speed, and the angle of the free
surface of the heap is recorded.  The value of $t_w$ and of the
maximum stability angle are determined from the last image before
the onset of the avalanche. Other experiments are then run from
step 2.
\end{enumerate}

 We have probed a large range of waiting times starting from $t_w=10$~s to
$t_w=3$~days. Experiments with different waiting times are
performed in random order.

In the case of high humidity or large waiting time, the granular
heap can be so cohesive that the heap can clump. In this case, it
is not possible to obtain a flat surface when the heap breaks. In
the following work, we have restricted our study to granular heaps
with low cohesion forces so that a flat surface can be obtained.

\section{Humidity induced aging}

\begin{figure}[htbp]
\includegraphics[width=7cm]{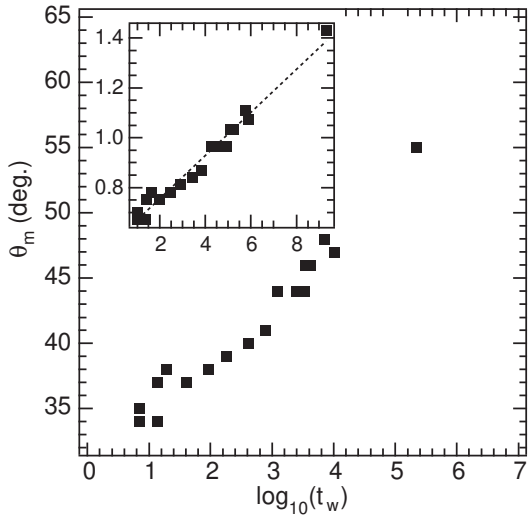}
\caption{The maximum stability angle $\theta_m$ as a function of
$\log t_w$ for a sample of glass beads with a diameter between 200
and 250~$\mu$m. In these experiments the value of the reference
angle is $\theta_{ref}=28^{\textrm o}$ and the relative humidity
is $RH\%=43\%$. In the insert, $\tan\theta_m$ is plotted as a
function of $\log t_w/\cos\theta_m$. The dashed line is the best
linear fit of the data. } \label{aging_fig}
\end{figure}
We first studied the influence of the resting time $t_w$ on sample
A in an atmosphere of relative humidity $RH=43\%$. The avalanche
angle $\theta_m$ is plotted as a function of $t_w$ in figure
\ref{aging_fig}. The data clearly show an increase of the
avalanche angle with the waiting time. This variation is large;
the maximum stability angle is about $30^{\textrm o}$ for a short
waiting time and rises up to $55^{\textrm o}$ after 3 waiting
days. Even for longer waiting times, no saturation of the maximum
stability angle was observed.

As a first approach, we may compare this temporal behaviour to the
aging properties of the static friction coefficient in normal
solid-to-friction. In a number of materials, the static friction
coefficient $\mu$ is found to increase logarithmically with the
waiting time during which the two solids have remained at rest
with respect to each other:~$\mu=a+b\log t_w$ . Following the law
of Coulomb for the avalanche angle of the granular heap, one may
expect a slow temporal growth for $\tan\theta_m$, for instance, as
$\tan\theta_m=a+b\log(t_w)$. However, this crude analogy is not
adequate to describe the aging behaviour observed here, since it
leads to a saturation value of the maximum stability angle equal
to $90^{\textrm o}$. Indeed, in a purely frictional granular
material, any negative normal stress applied on the surface of the
heap should destroy the heap. This was not observed in our
experiments. With large humidities ($RH\%>80\%$) and long waiting
times, we observed a clumping of the granular material. In such
situation the surface of the heap reaches $\theta>90^{\textrm o}$.
Clearly, cohesion forces retain
 the grains of the material together so that  the heap can
stay stuck on the upper part of the drum due to this cohesive
stress.

In order to take into account these adhesive forces, one may use a
phenomenological extension of the Coulomb analysis. An equivalent
formulation is to introduce the Mohr-Coulomb criterium, which
relates the normal stress $\sigma$ across some plane interior to
the sandpile to the maximum tangential stress $\tau$ possible
without failure of the pile. In the absence of cohesive stress, a
Mohr-Coulomb criterion for a granular medium is: $\tau=\mu\sigma$.
A cohesive stress can be taken into account by a phenomenological
criterion \cite{Nedderman}:
\begin{equation}\label{modMC}
\tau=c+\mu\sigma_n  \end{equation} where $\sigma_n$ is the normal
stress introduced by external fields, and $c$ the normal stress
due to attractive interaction between grains . The
maximum stability angle may then be obtained by considering the
equilibrium of a layer of thickness $D$ along the surface of the
heap. The modified Mohr-Coulomb criterium (\ref{modMC}) leads to
a failure when:
$$ \rho g D \sin \theta = c + \mu \rho g D \cos \theta$$
\begin{equation}\label{tant}
\tan\theta=\mu+\frac{c}{\rho g D \cos\theta}
\end{equation}
where $\rho$ is the specific weight of the pile. This
phenomenological description of the effect of cohesion forces on
the stability of the sand pile explains the fact that the free
surface of the pile can reach a vertical position. This occurs
when the cohesive stress $c$ is strong enough to sustain the
weight $\rho g D$ of the layer. Halsey et al.  have conducted this
analysis in more detail, and have predicted that in presence of
cohesion forces the failure of the heap should occur at its bottom
\cite{Halsey98}. Also, the dependency of the avalanche angle with
cohesion forces described in (\ref{tant}) has been probed recently
by Forsyth et al. \cite{Forsyth2001}, who have used well
controlled magnetic adhesion forces between grains.

In the insert of figure \ref{aging_fig}
we have plotted the tangent of $\theta_m$ as a function of $\log
t_w/\cos \theta_m$.  The data are
correctly fitted by the straight line such as:
\begin{equation}\label{equ_alpha}
\tan\theta_m=\tan\theta_0+\alpha\frac{\log t_w}{\cos \theta_m}
\end{equation}
where $\theta_0$ and $\alpha$ are two adjustable coefficients.
According to equation \ref{tant}, this result means that there is
a cohesion stress in the granular media which increases slowly in
time according to: $c(t_w)=c_0\log(t_w)$. In the following, we
characterize the influence of various parameters on the aging
behaviour by plotting the experimental data in this
representation. We
 focus mainly on the coefficient $\alpha$, which measures the the
 amplitude of the aging behaviour of friction in the granular medium.

\section{Parameter influencing the aging}

\subsection{The wear}

We have studied the influence of the preparation of the sample of
glass beads on their aging behaviour. For this purpose we have
taken $2.5$~kg of glass beads directly as bought from Matrasur,
and sifted them to calibrate their diameter in the 125-160~$\mu$m
range (sample B). We then have performed a first set of
measurements of the aging coefficient $\alpha$ of this sample.
After this first measurement, we have worn down the beads by
rotating the drum at a higher speed (60~rpm) during a number of
revolutions of the drum $N_c=N_1$.
\begin{figure}[htbp]
\includegraphics[width=7cm]{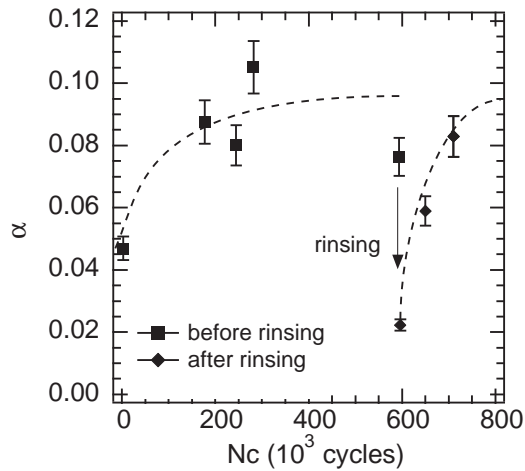}
\caption{Influence of the wear on the amplitude $\alpha$ of the
aging behaviour. $N_{c}$ is the number of revolution of the drum
after which aging is measured. An arrow indicates that after this
experiment, the beads have been rinsed as explained in the text.
The dashed line are only eye-guides.} \label{usure_fig}
\end{figure}
We then have produced a set of avalanches to measure the new value
of the aging coefficient. We then have worn down the beads again
during $N_2$ revolutions. The total number of revolutions of the
beads is now $N_c=N_1+N_2$. The evolution of the aging coefficient
$\alpha$ as a function of the total numbers of revolution $N_c$ is
plotted on figure \ref{usure_fig}. This curve shows a slow
increase of $\alpha$ with $N_c$ and a saturation of the aging
coefficient.

After these five measurements, the glass beads were taken out of
the drum and rinsed in distilled water several times. After the
first rinse,  the remaining water was turbid. We rinsed the beads
until the rinsing water is no longer turbid. This suppresses all
the small particles with a diameter typically smaller than
1~$\mu$m. The glass beads were dried, and the amplitude of aging
measured again. On figure \ref{usure_fig}, the rinsing operation
is symbolised by the vertical arrow. Just after rinsing, the aging
coefficient has a very small value, even lower than the one
obtained for the glass beads as coming from the supplier. We then
have worn down the sample again, and we observed that the aging
coefficient increases and comes back to the values obtained before
the rinsing operation.

Therefore, the amplitude of the aging of the beads depends on
their wear, but worn-down beads can be regenerated by rinsing.
This experiment reveals that the effect of wear is mainly due to
the production of glass dust, which plays an important role in the
aging phenomenon. Also, the effect of wear is important for
"clean" beads but tends to saturate after a certain number of
revolutions of the drum. As a result, we subsequently have used
worn-down beads to study the influence of other parameters on the
aging behaviour of the beads.

\subsection{Influence of the reference angle}

\begin{figure}[htbp]
\includegraphics[width=7cm]{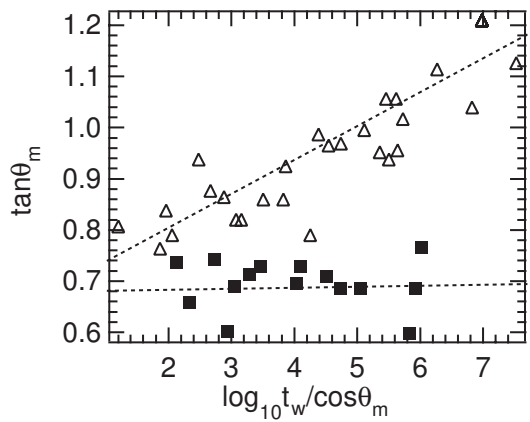}
\caption{Temporal evolution of the maximum stability angle for
different value of the reference angle $\theta_{ref}$ at which the
granular heap remains at rest. Filled squares :
$\theta_{ref}=0^{\textrm o}$ ; open triangles :
$\theta_{ref}=28^{\textrm o}$ } \label{angleref_fig}
\end{figure}
We also have studied the influence of the reference angle
$\theta_{ref}$ on the aging behaviour of friction in glass beads.
Figure \ref{angleref_fig} compares the evolution of the avalanche
angle for a sample of glass beads which ages with a horizontal
free surface ($\theta_{ref}=0^{\textrm o}$) and a sample of glass
beads which ages at a reference angle close to the first avalanche
angle measured at short time : $\theta_{ref}=28^{\textrm o}$. The
relative humidity in this experiment is $RH=43\%$. The beads' size
is the same as in the previous experiments (sample B). One can see
in figure \ref{angleref_fig} that  aging is much smaller for beads
which rest horizontally than for beads that rest at an angle close
to the avalanche angle.

We believe that this effect is due to very small displacements
occurring in the bulk of the heap when it is raised from an
horizontal position to the avalanche angle. The effect of these
small displacements is to break cohesive contacts between grains.
In their study of granular friction, Nasuno et al. \cite{Nasuno97}
report that in a stick-slip regime, some very  small displacements
(creep) precede the rapid events.
 The same precursors also have been reported in solid
friction by Baumberger et al. \cite{Baumberger95}. Those
precursors change dramatically the history of the adhesive
contacts by breaking old ones and reforming new ones. To induce
such changes, their amplitude has to overcome the range of the
cohesive forces. In our experiment such precursors are much more
numerous and important when the heap is tilted from an horizontal
position up to the avalanche angle, which explains the weak aging
effect obtained in this case. We will discuss in the last section
the expected range of the cohesion force between the grains. This
range must be small because the precursors obtained in solid
friction or in granular friction are about 1~$\mu$m at most.

\subsection{Influence of the surrounding humidity}

\begin{figure}[htbp]
\centering
\includegraphics[width=7cm]{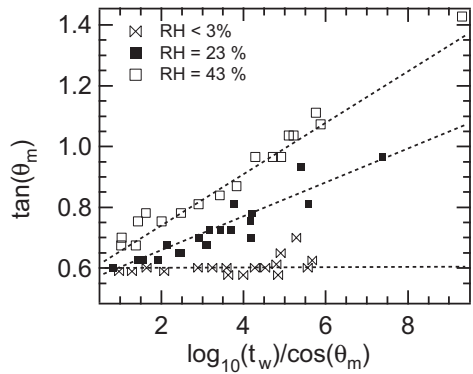}
\caption{Time evolution of the maximum stability angle
    $\theta_m$ of a $200$~$\mu$m glass beads heap for various relative humidity.}
    \label{thetam_fig}
\end{figure}
When the wear of the beads and  the reference angle are
conveniently chosen, the amplitude of the aging behaviour is
reproducible and the influence of the surrounding humidity can be
studied. The influence of the humidity $p_{vap}/p_{sat}$ on the
aging coefficient of a sample A of glass beads is plotted on
figure \ref{thetam_fig}. The main results are: i) that no aging is
observed in the dry case. That means that the  aging behaviour is
not due to any slow compaction effect of the heap(\cite{Nowak98}),
and ii) that the magnitude of the aging effect increases
dramatically with humidity.

At this point, an important question concerns the physical origin
of the slow increase of cohesion forces in an humid atmosphere
revealed by the aging phenomenon. In order to check whether the
influence of the water vapor is physical or chemical, we studied
the influence of an other condensable vapor : ethanol.

\section{Aging in an atmosphere of ethanol}

\subsection{Control of the partial pressure of ethanol}

Unlike the case of water, there is no tabulated value of the
influence of a  salt dissolved in liquid ethanol on its
equilibrium vapor pressure. Therefore, we first have designed a
method to control the partial pressure of ethanol vapor, based on
the one used to control water humidity.

The experimental setup is described in figure \ref{ethanol_fig}.
We measured the absolute pressure over a solution of ethanol
saturated with different salts  at different temperatures. The
solution of ethanol is placed in a bath with a controlled
temperature. The sensors then where thermalised at the temperature
of the bath to prevent any condensation of ethanol vapor.
\begin{figure}[htbp]
\includegraphics[width=7cm]{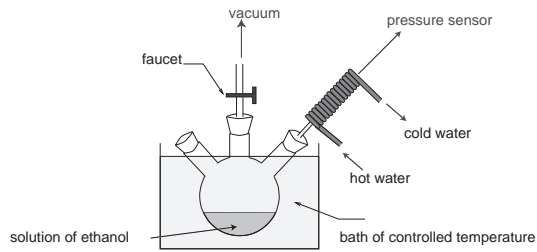}
\caption{Shematic representation of the experimental setup used to
calibrate the influence of an inorganic salt on the equilibrium
pressure of ethanol vapor in the surrounding atmosphere.}
\label{montagevap_fig}
\end{figure}
Because ethanol is a polar liquid, it can dissolve some inorganic
salts. In this study, we used MgCl$_2$, LiCl and LiBr. As for the
water solution, this modifies the chemical potential of ethanol in
the liquid phase and, thus, the pressure of the vapor in
equilibrium with this solution. Indeed, the partial pressure of
ethanol strongly depends on the temperature \cite{Reid}, but we
observed that the ratio of the partial pressure of the vapor at a
given temperature, divided by the tabulated saturating pressure,
does not really depend on the temperature. Typically, we obtained
$p_{vap}/p_{sat}=0.84$ for MgCl$_2$, $p_{vap}/p_{sat}=0.44$ for
LiCl, and $p_{vap}/p_{sat}=0.16$ for LiBr.

\begin{figure}[htbp]
\includegraphics[width=7cm]{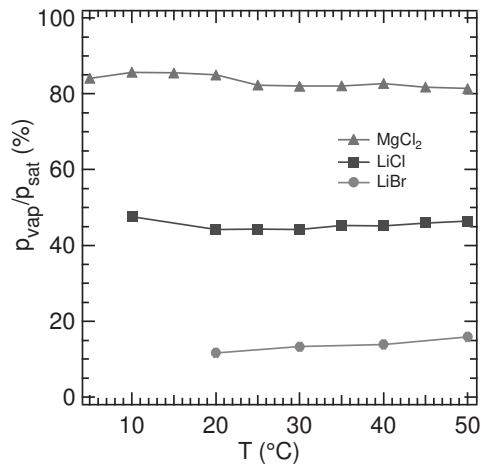}
\caption{Evolution of the equilibrium humidity of ethanol fro
different saturated solutions as a function of temperature. The
humidity is here defined as the ratio $p_{vap}/p_{sat}$ of the
equilibrium pressure $p_{vap}$ of ethanol vapor above the
saturated solution to the saturating vapor pressure $p_{sat}$ of
ethanol vapor above pure liquid ethanol.} \label{RHethanol_fig}
\end{figure}

\subsection{Aging in ethanol atmosphere}

\begin{figure}[htbp]
\includegraphics[width=7cm]{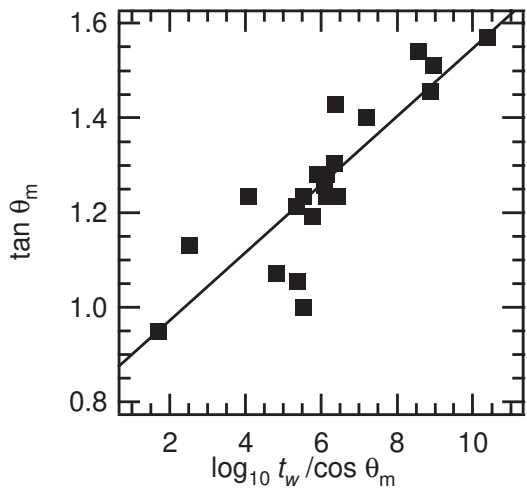}
\caption{$\tan\theta_m$ as a function of $\log t_w/\cos\theta_m$
for a sample of glass beads with a diameter comprise between 0 and
50~$\mu$m. The temperature is not controlled in this experiment. A
vapor pressure $p_{vap}/p_{sat}=0.84$ of ethanol is present in the
atmosphere of the box. The straight line is the best linear fit of
the data.} \label{ethanol_fig}
\end{figure}
In  an atmosphere of ethanol, we studied the evolution of the
first avalanche angle of glass beads with their resting time. We
have not found any aging behaviour for glass beads of $200$~$\mu$m
diameter. Since the surface tension of ethanol is lower than the
surface tension of water, we checked the effect of ethanol on a
sample of smaller glass beads. Figure \ref{ethanol_fig} shows that
aging can be observed for glass beads with a diameter comprised
between 0 and 50~$\mu$m. On this plot, the data are more dispersed
than for water vapor because temperature was not controlled in
these experiments  for safety reasons.

\section{Discussion}

These experiments on the aging behaviour of friction in a granular
heap of glass beads show that this aging is mainly due to a slow
increase in time of  cohesion forces inside the heap. As an order
of magnitude, we find that the avalanche angle of a heap of
$200$~$\mu$m diameter glass beads may increase from $35^{\textrm
o}$ to $55^{\textrm o}$ in two days in an ambient atmosphere of
$40\%$ relative humidity.

We have shown that this aging phenomenon is significantly
increased when the beads are worn down so as to contain a certain
amount of glass dust. This can be understood by the fact that
glass dust increases the specific area of the granular medium and,
thus, the total amount of liquid which can condense in it under
otherwise fixed conditions. Also, the aging behaviour is very
sensitive to the tangential stress applied to the medium when it
is at rest. Similar effects have been observed in usual solid
friction; they are due to small displacements that occur in the
medium when it tangentially is loaded and whose effect is to break
old contacts and to create new ones. This effect is particularly
important in our experiment. We do not know if it is an intrinsic
property of granular friction, or if it comes from the fact that
our granular medium is loaded only by its own weight. Finally, we
have not found any saturation of this temporal evolution of the
avalanche angle. We attribute this
 to the fact that when cohesion effects become important,
clumping and cohesive structures appear in the bulk of the heap,
and our experimental setup is not adapted to the measurement of a
maximum stability angle.

A very important parameter influencing aging is the humidity
content of the atmosphere. Bocquet et al. \cite{Bocquet98a} have
proposed a physical mechanism for this humidity induced aging. It
is based on  the slow increase of the cohesion force in a granular
medium due to the capillary condensation of small liquid bridges
between the grains. It has been shown both experimentally
\cite{Barabasi99} and theoretically \cite{Halsey98} that the
capillary forces created by liquid bridges can increase very
significantly the maximum stability angle of a granular medium.
Hornbaker et al. \cite{Hornbaker97} have also shown by adding
mineral oil to polystyrene beads - a chemically inert system -
that the capillary forces induced by liquid bridges are strong
enough to result in global clumping of the granular heap. In these
approach, time does not appear as a parameter since the liquid is
directly added to the granular medium. However, in the presence of
a condensable vapor, the liquid bridges could condense in the
medium with a slow kinetics. Direct measurements of capillary
condensation of a liquid bridge between solid surfaces performed
with surface forces apparatus \cite{Crassous94,Christenson87a}
show that a thin film of vapor confined between wetting surfaces
can be highly metastable, whereas the stable state is a liquid
bridge. Restagno et al. \cite{Restagno2000} have evaluated the
energy barrier for the condensation of such liquid bridges, and
have shown that the activation energy grows like the power $3/2$
of the distance between the solid surfaces, and may become very
large. Bocquet et al. \cite{Bocquet98a} then suggest a mechanism
of thermally activated capillary condensation of liquid bridges
between random asperities of the contact between grains. Assuming
a random distribution of the asperities, they obtain a large
distribution of activation energies, which result in a slow
increase in time of the capillary force between grains.

The  experiments reported here do not allow us to determine if the
increase of cohesion forces in the heap of glass beads in a humid
atmosphere is due to a purely physical mechanism, as proposed by
Bocquet et al, or a chemical mechanism, such as the sintering of
glass beads due to a chemical reaction between of silica and
water. We have observed an aging phenomenon in ethanol atmosphere,
which is much less reactive with glass than water. However, aging
in this case is observed only with beads with a diameter
significantly smaller than the one used with water. In order to
compare quantitatively the two systems, one  can try to estimate
by a dimensional argument the effect of capillary forces in the
bulk of the heap. The capillary force between two beads of radius
$r$ is $2\pi \gamma_{LV}r$, and the number of beads contacts
across a plane in the medium scales as $1/r^2$. Therefore, the
cohesive stress due to capillary force should scale as
$\gamma_{LV}/r$. On the other hand, the bulk density of the
granular medium does not vary much with the size of the beads.
Thus, assuming that the kinetics of condensation of liquid bridges
is the same in an ethanol or a water atmosphere, one would expect
similar effects for a ratio of beads radii of
$\gamma_{water}/\gamma_{ethanol}=2,8$. In our experiment, the
beads radii ratio in sample A and C is more than $5$, and aging is
observed in an ethanol atmosphere for a relative humidity much
higher than in a water atmosphere. Therefore, it is not possible
to conclude that capillary forces only contribute to the aging
phenomenon. In order to test more precisely the physical origin
and the time dependency of the adhesion force between beads, this
adhesion force should be measured directly between controlled
surfaces and in a controlled atmosphere. The surface forces
apparatus, which  directly measures the adhesion force between two
macroscopic surfaces, could be an ideal way to do these studies,
since it has been already used to study the kinetics of growing of
a liquid bridge of capillary condensation between ideal surfaces
\cite{Kohonen99}.

\acknowledgments We specially thank N. Taberlet for his
experimental help. It is a pleasure to thank L. Bocquet, J.
Crassous and N. Fraysse for discussions. We are grateful to F.
Vittoz and A. Gourinel for their technical help. F.R. thanks Nancy
Park for her careful reading of the manuscript.

\end{document}